\documentstyle[11pt,newpasp,twoside,epsfig]{article}
\def\edcomment#1{\iffalse\marginpar{\raggedright\sl#1\/}\else\relax\fi}
\reversemarginpar

\begin{document}
\title{Atomic Processes in Planetary Nebulae}
 \author{Sultana N. Nahar}
\affil{Department of Astronomy, The Ohio State University, 
Columbus, OH 43210, USA}

\begin{abstract}
A hot central star illuminating the surrounding ionized H II region 
usually produces very rich atomic spectra resulting from basic atomic 
processes: photoionization, electron-ion recombination, bound-bound 
radiative transitions, and collisional excitation of ions. Precise 
diagnostics of nebular spectra depend on accurate atomic parameters 
for these processes. Latest developments in theoretical computations 
are described, especially under two international collaborations 
known as the Opacity Project (OP) and the Iron Project (IP), that have 
yielded accurate and large-scale data for photoionization cross 
sections, transition probabilities, and collision strengths for electron 
impact excitation of most astrophysically abundant ions. As an extension 
of the two projects, a self-consistent and unified theoretical treatment 
of photoionization and electron-ion recombination has been
developed where both the radiative and the dielectronic recombination
processes are considered in an unified manner. Results from the Ohio 
State atomic-astrophysics group, and from the OP and IP collaborations, 
are presented. A description of the electronic web-interactive database, 
TIPTOPBASE, with the OP and the IP data, and a compilation of 
recommended data for effective collision strengths, is given. 
\end{abstract}

\section{Introduction}

A planetary nebula may be thought of as an `astrophysical 
laboratory' of atomic emission-line spectra. The spectra enable 
determination of its temperature, density, and abundances of elements. 
The precise spectral analyis require accurate atomic parameters for the 
radiative are collisional processes in the nebular plasma. The four 
basic atomic processes that dominant the nebular plasma are:

\noindent
i) Radiative bound-bound transition for excitation or de-excitation: 

$$X^{+Z} + h\nu \rightleftharpoons X^{+Z*},$$

\noindent
where the ion, $X$, is of charge, $Z$.

\noindent
ii) Photoionization (PI) by absorption of a photon:

$$X^{+Z} + h\nu \rightleftharpoons X^{+Z+1} + \epsilon,$$

\noindent
The inverse process is the electron-ion radiative recombination (RR).

\noindent
iii) Autoionization (AI) and dielectronic recombination (DR):

$$e + X^{+Z} \rightarrow (X^{+Z-1})^{**} \rightarrow \left\{ \begin{array}{ll}
e + X^{+Z} & \mbox{AI} \\ X^{+Z-1} + h\nu & \mbox{DR} \end{array}
\right. $$

\noindent
where the intermediate doubly excited state, autoionizing state, introduces 
resonances in the atomic processes. The inverse process of DR is 
photoionization via the autoionizing state.

\noindent
iv) The collisional process of electron-impact excitation (EIE)

$$e + X^{+Z} \rightarrow e' + X^{+Z*}*$$

\noindent
is one of the primary processes for spectral formation in astrophysical 
plasmas.

Each process needs to be treated separately to obtain the relevant 
atomic parameters. Sample results for each process, obtained mainly under 
the OP (1995, 1996) and the IP (1993) for accurate atomic data and 
stellar opacities will be presented later. Astrophysical model 
applications, such as for plasma opacities, need a huge amount of atomic 
data as these deal with a large number of atomic levels over wide 
energy ranges. Consistent sets of atomic parameters can be obtained if 
the atom or the ion is described by the same wavefunction expansion in 
each process. This also reduces the uncertainty in applications 
involving different processes and approximations. The Close Coupling (CC) 
R-matrix mehtodology employed by the OP and IP enable the computation of 
such self-consistent sets of atomic parameters. 

The large amount of atomic data and opacities obtained under the OP and 
the IP are available electronically through the existing database, TOPbase 
(Cunto et al. 1993), and through its planned extension TIPTOPBASE (C. 
Mendoza and the OP/IP team).

\section{Theory}

All calculations for the various atomic parameters of the dominant atomic
processes are carried out using the accurate and powerful R-matrix method 
in the close-coupling approximation (e.g. Burke \& Robb 1975, Seaton
1987, Berrington et al. 1987, Berrington et al. 1995). 
The total wavefunction for a (N+1) electron system in the CC 
approximation is described as:
\begin{equation}
\Psi_E(e+ion) = A \sum_i^N \chi_i(ion)\theta_i + \sum_{j} c_j \Phi_j(e+ion),
\end{equation}
\noindent
where $\chi_i$ is the target ion or core wavefunction in a specific state
$S_iL_i\pi_i$ or level $J_i\pi_i$, $\theta_i$ is the wavefunction of the 
interacting (N+1)th electron in a channel labeled as
$S_iL_i(J_i)\pi_i \ k_{i}^{2}\ell_i(SL\pi~or~ \ J\pi)$; $k_{i}^{2}$ is the
incident kinetic energy. 
$\Phi_j$ is the correlation functions of (e+ion) system that compensates
the orthogonality condition and short range correlation interations.
The complex resonant structures in photoionization, recombination, and in
electron impact excitation are included through channel couplings.

Relativistic effects are included through Breit-Pauli approximation in
intermediate coupling. The (N+1)-electron Hamiltonian in the Breit-Pauli
approximation, as adopted in the Iron Project, is
\begin{equation}
H_{N+1}^{\rm BP}=H^{NR}_{N+1}+H_{N+1}^{\rm mass} + H_{N+1}^{\rm Dar}
+ H_{N+1}^{\rm so},
\end{equation}
\noindent
where non-relativisitc Hamiltonian is
\begin{equation}
H^{NR}_{N+1} = \sum_{i=1}\sp{N+1}\left\{-\nabla_i\sp 2 - \frac{2Z}{r_i}
        + \sum_{j>i}\sp{N+1} \frac{2}{r_{ij}}\right\} . 
\end{equation}
\noindent
$H_{N+1}^{\rm mass}$ is the mass correction term, $H_{N+1}^{\rm Dar}$ is
the Darwin term, $H_{N+1}^{\rm so}$ is the spin-orbit interaction term.
Spin-orbit interaction splits the LS terms into fine-structure levels
labeled by $J\pi$ where $J$ is the total angular momentum. 
Solutions of the Schrodinger equation, $H^{BP}_{N+1}\mit\Psi = E\mit\Psi$
which becomes a set of coupled equations with the CC expansion, give the 
bound wavefunctions, $\Psi_B$, for negative energies, E $<$ 0, 
and continuun wavefunction, $\Psi_F$, for positive energies, E $\geq$ 0.

The transition matrix elements for various atomic processes are: \\
$<\Psi_B || {\bf D} || \Psi_{F}>$ for photoionization and recombination, \\
$<\Psi_B || {\bf D} || \Psi_{B'}>$ for oscillator strength, \\
$<\Psi_F | H(e + ion) | \Psi_{F'}> $ for electron impact excitation, \\
where {\bf D} is the dipole operator.

The transition matrix element with the dipole operator can be reduced
to the generalized line strength defined, in either length or velocity
form, as
\begin{equation}
S_{\rm L}=
 \left|\left\langle{\mit\Psi}_f
 \vert\sum_{j=1}^{N+1} r_j\vert
 {\mit\Psi}_i\right\rangle\right|^2 \label{eq:SLe},
~~~S_{\rm V}=\omega^{-2}
 \left|\left\langle{\mit\Psi}_f
 \vert\sum_{j=1}^{N+1} \frac{\partial}{\partial r_j}\vert
 {\mit\Psi}_i\right\rangle\right|^2. \label{eq:SVe}
\end{equation}
where $\omega$ is the incident photon energy in Rydberg units, and
$\mit\Psi_i$ and $\mit\Psi_f$ are the wave functions representing the
initial and final states, respectively.

The oscillator strength $f_{ij}$ and the transition probability $A_{ji}$
for the bound-bound transition can be obtained from $S$ as
\begin{equation}
f_{ij} = {E_{ji}\over {3g_i}}S, 
~~~A_{ji}(a.u.) = {1\over 2}\alpha^3{g_i\over g_j}E_{ji}^2f_{ij},
\end{equation}
where $E_{ji}$ is the transition energy, $\alpha$ is the fine structure 
constant, and $g_i$, $g_j$ are the statistical weight factors of the 
initial and final states. The lifetime, $\tau_j$, of a level $j$ can be 
obtained from the transition probabilities decaying to the lower levels 
as $\tau_j = (\sum_i A_{ji}(s^{-1}))^{-1}$, where $A_{ji}(s^{-1}) = 
{A_{ji}(a.u.)\over \tau_0}$ and $\tau_0 = 2.4191\times 10^{-17}$s.

The photoionization cross section ($\sigma_{PI}$) is proportional to 
the generalized line strength ($S$),
\begin{equation}
\sigma_{PI} = {4\pi \over 3c}{1\over g_i}\omega S.
\end{equation}

The calculations of total electron-ion recombination employs a new
unified treatment (Nahar \& Pradhan 1994,1995). The method
considers the infinite number of recombining states and incorporates
the radiative and dielectronic recombinations, RR and DR, in a unified
manner. The contributions to recombination from states with $n \leq$ 
10 are obtained from $\sigma_{PI}$ using principle of detailed balance,
while the resonant contributions from states with 10 $< n \leq ~ 
\infty$ are obtained from an extension of the DR theory of Bell \& Seaton 
(1985).

The recombination cross section, $\sigma_{RC}$, is related to 
$\sigma_{PI}$ through the principle of detailed balance,
\begin{equation}
\sigma_{RC} = \sigma_{PI}{g_i\over g_j}{h^2\omega^2\over 4\pi^2m^2c^2v^2}.
\end{equation}
The recombination rate coefficient, $\alpha_{RC}$, is obtained as
\begin{equation}
\alpha_{RC}(T) = \int_0^{\infty}{vf(v)\sigma_{RC}dv},
\end{equation}
where $f(v)$ is the Maxwellian velocity distribution function. The total 
$\alpha_{RC}$ is obtained from contributions from infinite number of 
recombined states.

The collision strength for transition by electron impact excitation from 
the initial state of the target ion $S_iL_i$ to the final state $S_jL_j$ 
is given by
\begin{equation}
\Omega(S_iL_i-S_jL_j) = {1\over 2}\sum_{SL\pi}\sum_{l_il_j}(2S+1)
(2L+1)|{\bf S}^{SL\pi}(S_iL_il_i - S_jL_jl_j)|^2,
\end{equation}
where ${\bf S}$ is the scattering matrix. The effective collision strength 
or the Maxwellian averaged collision strength can be obtained as
\begin{equation}
\Upsilon(T)=\int_0^{\infty} \Omega_{ij}(\epsilon_j)e^{-\epsilon_j \over kT}
d(\epsilon_j/kT).
\end{equation}
which give the excitation rate coefficient,
$q_{ij}(T)=(8.63\times 10^{-6}/ \omega_iT^{1/2}) e^{-E_{ij}/kT} \Upsilon (T)$
in $~cm^3s^{-1}$.
where $T$ is in K, $E_{ij}=E_j-E_i$, $E_i<E_j$ are in Rydbergs (1/kT = 
157885/T), and $j$ is the excited upper state.

\section{Results and Discussions}

Sample results for each atomic process are presented, e.g., for 
$f$-values, $\sigma_{PI}$, $\alpha_R(T)$, and $\Upsilon(T)$, in the 
following subsections.

\subsection{Bound-Bound transitions}

Due to fine structure, extensive sets of data for $f$- and $A$-values 
can be obtained for a large number of bound-bound transitions using the 
Breit-Pauli R-matrix method. In contrast to dipole allowed $LS$ multiplets, 
BPRM method includes relativistic effects, and consideration of both 
the dipole allowed and intercombination transitions. The fine structure 
energy levels are analysed with quantum defects to obtain the spectroscopic 
identification, $(C_t \ S_t \ L_t \ J_t~\pi_t n\ell)SL J \ \pi$ where
$(C_t \ S_t \ L_t \ J_t~\pi_t)$ denotes the core configuration, spin,
orbital, total angular momenta and parity, $nl$ is configuration of the
outer or valence electron, $SL$ are the possible spin and orbital angular
momenta, and $J\pi$ are the total angular momentum and parity
of the (N+1)-electron system. The fobidden quadrupole (E2) and magnetic 
dipole (M1) transitions are treated with atomic structure calculations 
using codes such as SUPERSTUCTURE (Eissner et al. 1974).

Accurate fine structure transition probabilities have been
obtained for a number of
ions such as: \\
Li-like: C IV, N V, O VI, F VII, Ne VIII, Na IX, Mg X, Al XI, Si XII,
S XIV, Ar XVI, Ca XVIII, Ti XX, Cr XXII, Ni XXVI \\ 
Fe ions: Fe V, Fe XVII, Fe XXI, Fe XXIII, Fe XXIV, Fe XXV  \\
Other ions: C II, C III, O IV, S II, Ar XIII

The first large-scale application of the BPRM method for a complex ion, 
Fe~V, resulted in 3865 fine structure energy levels, compared to 179 
observed, and about 1.46 million $f$- and $A-$values for dipole allowed 
and intercombination (E1) transitions and 362 forbidden (E2,M1)
transitions (Nahar et al. 2000). Table I presents a sample with complete 
spectroscopic identification of the transition array 
$3d^4(^4D)-~ 3d^3(^4F)4p(^5F^o)$ in Fe~V.

\begin{table}
\noindent{Table I: Transition probabilities of Fe V. g=2J+1. \\}
\normalsize
\begin{tabular}{llcccccc}
\hline
\noalign{\smallskip}
\multicolumn{1}{c}{$level_i$} &\multicolumn{1}{c}{$level_j$} &
$g_i$ & $g_j$ & $E_i(Ry)$ & $E_j(Ry)$ & $f$ & $A(sec^{-1})$ \\
 \noalign{\smallskip}
\hline
 \noalign{\smallskip}
 $3d^4(^5D)    $ & $3d^3(^4F)4p(^5F^o)$ &  1 &  3 &    5.5132 &   3.1644 &  2.154E-01 &  3.18E+09 \\
 $3d^4(^5D)    $ & $3d^3(^4F)4p(^5F^o)$ &  3 &  3 &    5.5119 &   3.1644 &  3.790E-04 &  1.68E+07 \\
 $3d^4(^5D)    $ & $3d^3(^4F)4p(^5F^o)$ &  3 &  5 &    5.5119 &   3.1496 &  1.358E-03 &  3.65E+07 \\
 $3d^4(^5D)    $ & $3d^3(^4F)4p(^5F^o)$ &  5 &  3 &    5.5094 &   3.1644 &  4.617E-02 &  3.40E+09 \\
 $3d^4(^5D)    $ & $3d^3(^4F)4p(^5F^o)$ &  5 &  5 &    5.5094 &   3.1496 &  5.967E-02 &  2.67E+09 \\
 $3d^4(^5D)    $ & $3d^3(^4F)4p(^5F^o)$ &  5 &  7 &    5.5094 &   3.1443 &  1.462E-02 &  4.69E+08 \\
 $3d^4(^5D)    $ & $3d^3(^4F)4p(^5F^o)$ &  7 &  5 &    5.5058 &   3.1496 &  6.895E-03 &  4.30E+08 \\
 $3d^4(^5D)    $ & $3d^3(^4F)4p(^5F^o)$ &  7 &  7 &    5.5058 &   3.1443 &  5.889E-02 &  2.64E+09 \\
 $3d^4(^5D)    $ & $3d^3(^4F)4p(^5F^o)$ &  9 &  7 &    5.5015 &   3.1443 &  1.966E-03 &  1.13E+08 \\
 $3d^4(^5D)    $ & $3d^3(^4F)4p(^5F^o)$ &  7 &  9 &    5.5058 &   3.1391 &  3.262E-02 &  1.14E+09 \\
 $3d^4(^5D)    $ & $3d^3(^4F)4p(^5F^o)$ &  9 &  9 &    5.5015 &   3.1391 &  5.139E-02 &  2.30E+09 \\
 $3d^4(^5D)    $ & $3d^3(^4F)4p(^5F^o)$ &  9 &  11 &   5.5015 &   3.1343 &  7.548E-02 &  2.78E+09 \\
 & & & & & & & \\
 $3d^4(^5D)    $ & $3d^3(^4F)4p(^5F^o)$ & 25 &  35 &   5.5055 &   3.1451 &  1.068E-01 &  3.42E+09 \\
\noalign{\smallskip}
\hline
\end{tabular}
\end{table}

\subsection{Photoionization and Recombination}

The photoionization cross sections, $\sigma_{PI}$, are calculated including
autoionizing resonances that can enhance the background cross sections
considerably. Fig.~1 shows the photoionization cross sections of the
gound states of Fe~I to Fe~V (Bautista \& Pradhan 1998). Extensive 
resonances dominate the cross sections for these complex ions. The 
enhancement in the backgound is up to three orders of magnitude for Fe I, 
over an order or magnitude for Fe II, and $\sim$ 50\% for Fe III. 

\begin{figure} %
\vspace*{-2.5cm}
\epsfig{figure=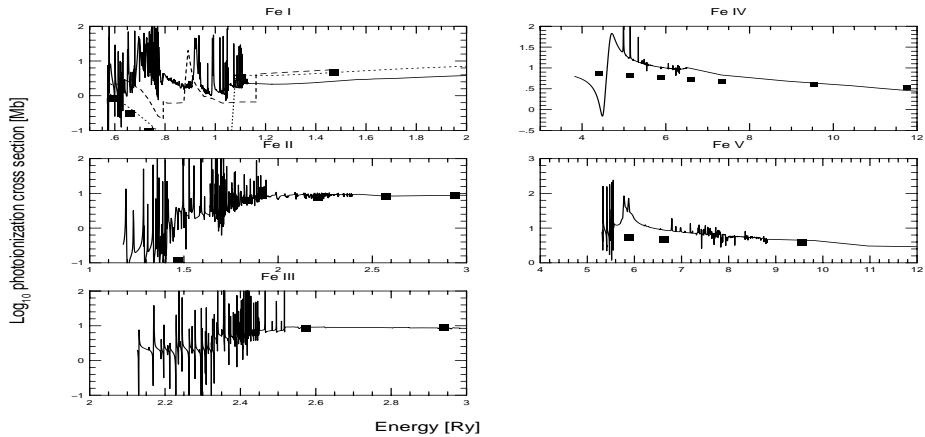,height=9.0cm,width=11.0cm}
\vspace*{-1.0cm}
\caption{Photoionization cross sections, $\sigma_{PI}$, of the ground 
states of Fe I - Fe V.}
\end{figure}

For simpler systems the accuracy of theoretical cross sections can be
tested against experimental data. Very precise measurements
of $\sigma_{PI}$ are now being carried out at a few places, such as
the Advanced Light Source at Berkeley (e.g. Covington et al. 2001),
the merged photon-ion beam set-up at Aahus University (Kjeldsen et al.
1999), and a synchroton based experiment at University of Paris-Sud 
(Wuilleumier et al., private communication). An example of the comparison 
of ground state photoionization cross sections of C II, with very good 
agreement with experiment, is shown in Fig. 2. Theoretical calculations 
include relativistic fine structure.

\begin{figure} %
\vspace*{-2cm}
\epsfig{figure=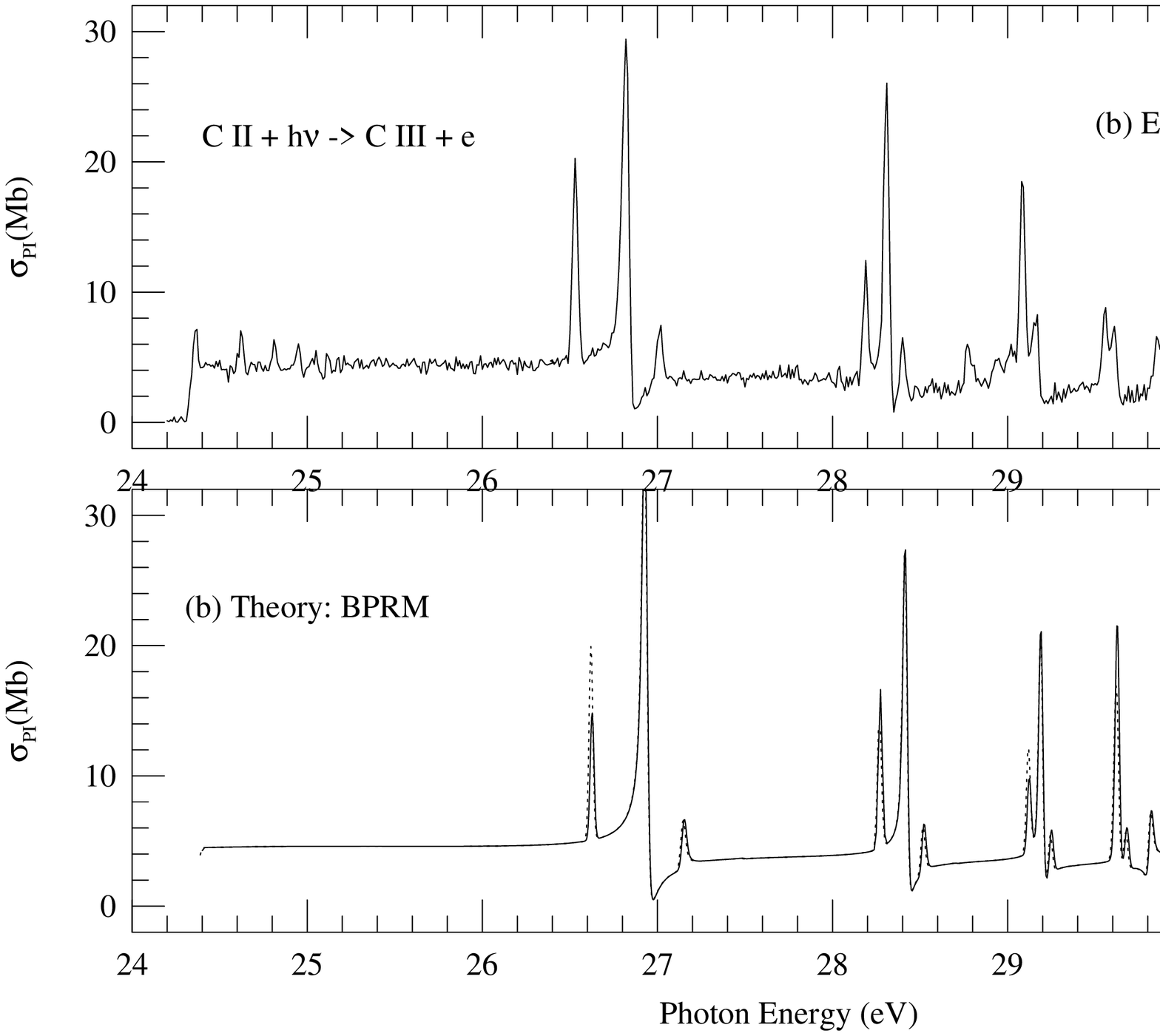,height=8.0cm,width=11.0cm}
\vspace*{-1cm}
\caption{Comparison of theoretical and experimental photoionization cross 
sections, $\sigma_{PI}$, of the ground state fine structure levels
$2s^22p (^2P^o_{1/2,3/2}$) of C II (Nahar 2002).}
\end{figure}

An example of the total unified recombination rate coefficient 
$\alpha_R(T)$ for O~III is presented in Fig. 3 (solid) (Nahar 1998). 
The earlier results are RR rates (dashed, Pequignot et al. 1991), 
low temperature DR rates (dotted, Nussbaumer \& Storey 1983), high 
temperature DR rates by Badnell \& Pindzolla (1990, short-dash-long-dash) 
and by Shull \& Steenberg (1982, dot-desh). Differences can be noticed 
between the present unified values and the sum of the earlier (RR+DR) 
results, especially in the low and intermediate temperature. Compared to 
the earlier Shull \& Steenberg results, autoionization into excited levels
in the high temperature region reduces the recombination rates significantly.

\begin{figure} %
\vspace*{-0.7cm}
\epsfig{figure=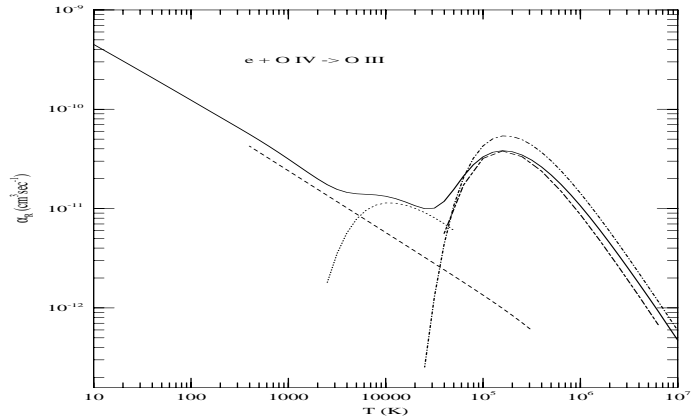,height=6.5cm,width=11.0cm}
\vspace*{-0.7cm}
\caption{Comparison of the unified total recombination rate coefficients 
($\alpha_R(T)$ (solid) with the previous calculations.}
\end{figure}

The unified treatment for the total recombination provides rates that are
valid over a wide range of temperatures for all practical purposes in
contrast to addition of RR and DR rates obtained using different 
approximations for different temperature ranges. The method enables 
obtaining self-consistent sets of photoionization and recombination cross
sections by using the identical wavefunction expansion for both the 
inverse processes. Following is the list of atoms and ions for which 
self-consistent sets of $\sigma_{PI}$ and $\alpha_R(T)$ have so far
been obtained for over 45 ions (e.g. Nahar \& Pradhan 1997, 
$http://www.astronomy.ohio-state.edu/\sim pradhan$):

\noindent
Carbon: C I, C II, C III, C IV, C V, C VI \\
Nitrogen: N I, N II, N II, N IV, N V, N VI, N VI \\
Oxygen: O I, O II, O III, O IV, O V, O VI, O VII, O VII \\
Iron: Fe I, Fe II, Fe III, Fe IV, Fe V, Fe XIII, Fe XVII,
Fe XXI, Fe XXIV, Fe XXV, Fe XXVI \\
C-like: F IV, Ne V, Na VI, Mg VII, Al VIII, Si IX, S XI, Ar XIII, Ca XV \\
Other ions: Si I, Si II, S II, S III, Ar V, Ca VII, Ni II 

\noindent
The data include the state specific recombination rates for hundreds of 
bound levels with n$\le$ 10 for each ion.

 The self-consistent sets of photoionization/recombination data include
new photoionization cross sections that are generally an improvement
over the OP data since more extensive and accurate eigenfunction
expansions are employed.

\subsection{Electron Impact Excitation}

 Reviews and compilations of available theoretical data sources are:
{\it An evaluated compilation of theoretical data sources for
electron-impact excitation of atomic ions}, (A.K. Pradhan \& J.W. 
Gallagher, Atomic Data and Nuclear Data Tables, 52, 227, 1992), and 
{\it Electron Collisions with Atomic Ions} (A.K. Pradhan \& H.L. Zhang, 
in LAND\"{O}LT-BORNSTEIN Volume {\it Atomic Collisions}, Ed. Y. Itikawa,
Springer-Verlag, in press).

A table of recommended data for effective collision strengths and A-values for
nebular ions is available on-line from 
www.astronomy.ohio-state.edu/$\sim$pradhan

The aim of the IP is to compute collisional data for the iron-peak 
elements in various ionization stages. Of the 50 publications in the 
"Atomic data from the IRON Project"  series in Astronomy and Astrophysics 
journal, most are on collisional strengths. At present the Iron Project 
website is maintained by Keith Butler at:
$www.usm.uni-muenchen.de/people/ip/iron-project.html$ 

An example of IP calculations for iron ions is the recent work on Fe~VI
(Chen \&  Pradhan 1999). Fig. 4 presents the detailed collision strength
$\Omega(E)$ showing extensive resonance structure. Previous calculations
neglecting resonances seriously underestimate the Maxwellian averaged
effective collision strength $\Upsilon(T)$ (Eq. 10) by several factors.

\begin{figure} %
\vspace*{-0.7cm}
\epsfig{figure=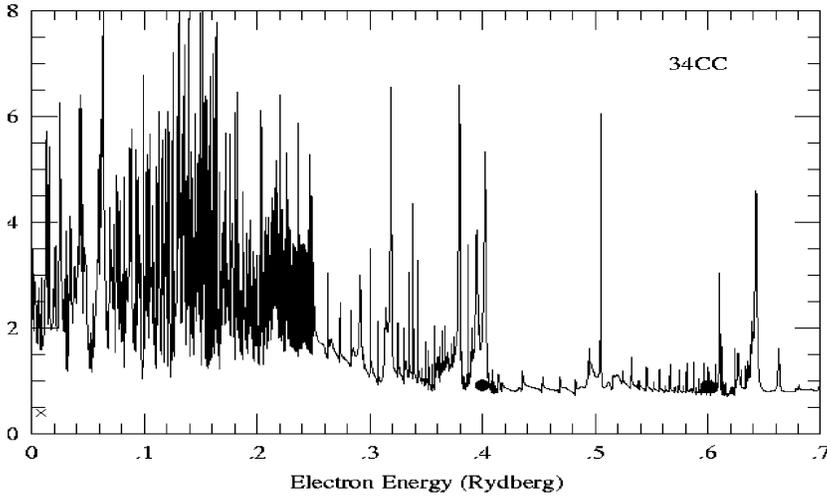,height=7.5cm,width=11.0cm}
\vspace*{-0.7cm}
\caption{Collision strength for the first transition in Fe~VI
$\Omega(^4F_{3/2}~-~^4F_{5/ 2})$ from the Iron Project CC R-matrix 
calculation (Chen \&  Pradhan 1999); previous results ($\bullet$ -
Nussbaumer \&  Storey 1978,  $\times$ - Garstang et al. 1978) neglect
resonances.}
\end{figure}

\section{TIPTOPBASE: atomic radiative and collisional data} 

The existing electronic archive for the OP data, TOPbase 
($http://heasarc.gsfc.nasa.gov$ at Goddard, NASA and
$http://vizier.u-strasbg.fr/OP.html$ at CDS) contain 
(i) photoionization cross sections of bound $LS$ terms for $\sim$ 200
ions with Z = 1 - 14, 16, 18, 20, 26,
(ii) energy levels and transition probabilities ($\sim 10^7~~f$-values),
(iii) monochromatic and Rosseland mean opacities (at CDS only)

The new database, TIPTOPbase (under development, C. Mendoza \& the OP/IP
team) will have radiative and 
collisional data from the OP and the IP. The data include:

(i) All data from TOPbase (and replacement of improved data),
(ii) collisional data for iron and iron peak elements, 
(iii) new elements through all ionization stages: P, Cl, K, as well as
selected ions such as Ni II and Ni III,
(iv) updated sets of radiative data from new CC calculations,
(v) "Tail" photoionization cross sections at high energies that include 
inner-shell ionization,
(vi) total and level specific recombination rate coefficients,
(vi) additional data for $f$-values including inner-shell excitations in
iron ions Fe~VIII - XIII  ("PLUS" data) calculated with SUPERSTRUCTURE,
(vii) radiative data ($\sigma_{PI}$ and $f$-values) for fine structure 
levels including relativistic effects,
(viii) on-line computational facilities for opacities and radiative
accelerations for user-specified mixtures of elements (``customized"
opacities and radiative forces) (see IP in Extended Abstracts).

\section{Conclusion}

The current status of large-scale ab initio close coupling R-matrix
calculations for radiative and collisional processes is reported.
The Iron Project Breit Pauli R-matrix radiative calculations include
large numbers of dipole allowed and intercombination transitions.
Self-consistent sets of atomic data for photoionization
and unified (electron-ion) recombination (including RR and DR)
are obtained, and should yield more accurate photoionization models.
Work is in progress for heavy ions of the iron group elements.

\acknowledgements{This work is supported partially by the U.S. 
National Science Foundation and NASA.}

\end{document}